# Are Mobilities in Hybrid Organic-Inorganic Halide Perovskites Actually 'High'?

Thomas M. Brenner[¶,1], David A. Egger[¶,1], Andrew M. Rappe[2], Leeor Kronik[1], Gary Hodes[1], David Cahen[1]

[1] *Department of Materials and Interfaces, Weizmann Institute of Science, Rehovoth, Israel 76100*

[2] *The Makineni Theoretical Laboratories, Department of Chemistry, University of Pennsylvania, Philadelphia, PA 19104–6323, USA*

[¶] These authors contributed equally

The outstanding performance of hybrid organic-inorganic perovskites (HOIPs) in photovoltaic devices is made possible by, among other things, outstanding semiconducting properties: long *real* charge-carrier diffusion lengths, $L$, of up to 5 and possibly even 10 μm, as well as a lifetime, $\tau$ of ~1 μs or more in single crystal and polycrystalline films.[1–9] Top electronic transport materials will have a high "μτ" product, the product of the charge carrier mobility, $\mu$, and lifetime. This is directly related to the diffusion length ($L=\sqrt{D\tau}$), where $D$ is the carrier diffusion coefficient given as $D=(\mu q/k_B T)$, where $q$ is the electron charge, $k_B$ is the Boltzmann constant, and $T$ is the absolute temperature. Long lifetimes, which imply slow recombination and low trapping probabilities, do not automatically imply high mobilities, which are limited by scattering.

Charge carrier mobilities in HOIPs are often described as "high", but this statement warrants some scrutiny. HOIP mobilities are often compared to those of charge carriers in organic semiconductors (see Table I) and are then indeed much higher.[10] But in our opinion, carrier mobilities in HOIPs need to be placed in the context of typical inorganic semiconductors, for several reasons: First, HOIPs exhibit a band structure resembling that of a good inorganic semiconductor, with the conduction and valence band dominated by the inorganic cations and anions, respectively.[11–14] This leads to small computed charge-carrier effective masses[11,13,14]: a reduced effective mass on the order of 0.1 electron mass, close to that of Si (0.08) or GaAs (0.03)[15], a value confirmed by magneto-absorption measurements.[16,17] Second, material disorder, which often lowers mobilities dramatically, is low: HOIPs exhibit sharp x-ray diffraction peaks,[18] small Urbach tail energy (~15 meV),[19,20] and low trap-state density (~$10^{10}$ cm$^{-3}$ in single crystals).[5–7] Despite these material properties, it is clearly seen in Table 1 that *mobilities in HOIPs are in fact rather modest*[10] - at least one order of magnitude lower in electron mobility (and at least several times lower in hole mobility) than those of Si, GaAs, and some other inorganic PV materials.

What could underlie these rather modest mobility values? The mobility is proportional to the carrier scattering time and inversely proportional to the effective mass.[15] If effective mass values of HOIPs are indeed on par with those of other inorganic semiconductors, then the mobility must be limited by scattering. Then again, the observation of long carrier lifetimes and an inverse power dependence of mobility on temperature (see below) suggests negligible impurity scattering at RT,[15] a fact which is consistent with the observation of slow carrier cooling at room temperature.[21] Therefore, we posit instead that an important hint for the origin of this phenomenon lies in the mechanical and vibrational properties of HOIPs. In particular, the bulk and Young's moduli of HOIPs are rather small, showing that these materials are relatively "soft",[22–27] with corresponding low speeds of sound.[23,26] Furthermore, given the B-X connectivity typical for $ABX_3$ perovskites, the mechanical properties of HOIPs are very similar to those of lead di-halides, indicating that it is the inorganic lead-halide framework that determines the mechanical properties.[26]

Together with the relatively large unit cells, the mechanical properties also result in low Debye temperatures and small (non-molecular) phonon activation energies.[23] Around RT, HOIPs can therefore be expected to exhibit structural fluctuations and dynamical disorder.[28] For methylammonium lead halide HOIPs ($MAPbX_3$, where X is a halogen atom), these will include also rotations of the entire MA unit around its axis and $PbX_6$ octahedral distortions, as confirmed by both molecular dynamics calculations and neutron scattering experiments.[29–31] Furthermore, it has been shown theoretically that these structural fluctuations are coupled to important changes in electronic structure.[29,31–33] One possibility, then, for a dominant charge-carrier scattering mechanism is relatively strong electron-phonon coupling. Indeed, several THz spectroscopy studies have indicated that mobilities show a $T^{-1.3}$ to $T^{-1.6}$ dependence between 150 K to RT,[34–37] interpreted as indicating phonon scattering because a dependence of $\sim T^{-3/2}$ is the fingerprint of electron-phonon scattering of charge carriers, specifically of deformation potential scattering by acoustic phonons.[15,34,36,37] Importantly, these results agree with temperature-dependent charge transport and Hall effect measurements of $MAPbBr_3$ single crystals.[38] If phonon scattering, rather than impurity scattering, dominates, this can explain the long lifetimes and modest mobilities.

The above explanation is not without difficulties. To reconcile the acoustic phonons with the sharp (close to laboratory diffractometer–limited) X-ray diffraction (XRD) line widths, notably for high order diffraction peaks (i.e., providing information on a large volume), found for high quality materials, the lattice vibrations need to have spatial coherence. This situation calls for additional studies, using high-resolution time-resolved XRD measurements, preferably under illumination, which could shed light on the role of coherent phonons.[39] In fact, based on agreement between measured and molecular dynamics-simulated pair distribution functions, it can be argued that the strong RT structural fluctuations of HOIPs are not in contradiction with sharp XRD data.[28,31,40] Furthermore, some theoretical reports predict that charge carrier mobilities under acoustic phonon scattering should be high ($10^3$ $cm^2/Vs$),[41,42] in conflict with the current experimental data (Table 1). Lastly, the fact that the mobility features significant temperature dependence needs to be reconciled with the fact that the theoretically-predicted Debye temperature[23] is much lower than the measured temperatures. The reasons for these discrepancies are currently far from being fully understood, but this implies that a deeper understanding of electron-phonon coupling of HOIPs is needed, likely on both the theoretical and experimental side.

Alternative scattering mechanisms for HOIPs are actively investigated. One possibility, compatible with the strong coupling between structural and electronic fluctuations mentioned above, which could also explain the observed carrier lifetimes, is conduction *via* polarons. Polaronic effects are invoked to explain low mobilities derived from ac Hall measurements,[43] as well as light-induced changes of the low frequency dielectric response.[44] Furthermore, light-generated carriers were suggested to couple to polar fluctuations in the crystal lattice, which is an indicator for polaron formation and could also play a role in the photostability of HOIPs.[45] Polar fluctuations in HOIPs can also locally support charge-carrier separation[46,47] or possibly suppress recombination through a Rashba effect,[48] which would contribute to the long carrier lifetime. However, the presence of polaronic effects would also increase the charge-carrier effective masses, which needs to be reconciled with the recent experimental findings of a small effective mass, mentioned above,[17] underlining the need for additional experimental studies of the effective masses in HOIPs, preferably under illumination. Furthermore, as mentioned above the experimentally determined temperature-dependence of the mobility has been of a $T^{-1.3}$ to $T^{-1.6}$ nature, which would need to be explained in terms of polaron-dominated scattering.

Finally, we note that while Pb-based HOIPs have the largest and most established data set, many compounds exist in the metal-halide perovskite family whose charge transport properties are as yet minimally explored. Carrier mobilities up to 2000 cm$^2$/Vs have been reported for some members of this family: MASnI$_3$ (e$^-$ mobility), CsSnI$_3$, and CsPbBr$_3$.[49,50] These findings have yet to be confirmed independently. However, if they stand they will surely illuminate the fundamental mechanisms determining mobility in this class of materials and may even indicate how mobilities might be improved in the materials discussed in this viewpoint.

In summary, in light of their highly favorable semiconducting properties, the mobilities of typical HOIP materials are actually not that high, especially when compared to inorganic semiconductors used as absorbers in high efficiency PV cells. If the semiconducting properties of HOIPs, especially the effective masses, are as favorable as the current literature reports them to be, then strong carrier scattering mechanisms must be active at RT. Pertinent literature data for the temperature-dependence of the mobilities suggest that these do not arise from impurities, but must be due to electron-lattice coupling. We have discussed that scattering due to acoustic-phonons, given experimental data on the temperature-dependence of the mobility, is currently a likely mechanism active at RT. Strong acoustic phonon scattering may originate from the soft mechanical properties of HOIPs. As this explanation is not straightforward to reconcile with XRD data and current theoretical predictions, we furthermore discussed the possibility of polaronic effects, which could also contribute to the charge-carrier lifetime of HOIPs. From this discussion, we conclude that determining the actual scattering mechanism(s) will require more (and more diverse) experimental data and theoretical models, as it may be influenced by a number of different effects. Such determination could provide great insight towards improving the charge-transport properties of HOIPs even further. We therefore hope that this short opinion piece will motivate further experimental and theoretical studies of this issue.

**Acknowledgements**
We are very grateful to Drs. M. Bonn, E. Canovas, V. Podzorov, S. Tretiak, and O. Yaffe for sharing preprints of their results with us. We thank Drs. I. Balberg A. Kahn, L. Leiserowitz, I. Lubomirsky, O.M. Stafsudd and X. Y. Zhu for illuminating discussions. LK and DAE were supported by a research grant from Dana and Yossie Hollander, in the framework of the Weizmann Institute of Science (WIS) Alternative sustainable Energy Initiative, AERI. TMB is an AERI postdoctoral fellow. DAE acknowledges financial support by the Austrian Science Fund (FWF):J3608−N20. GH and DC thank the Israel Ministry of Science for support. AMR acknowledges support from the Office of Naval research, under grant N00014-14-1-0761. DC holds the Sylvia and Rowland Schaefer Chair in Energy Research.

**Table 1**. Representative* charge-carrier mobilities for common PV materials at RT.

| Material | Mobility ($cm^2$/Vs) | |
|---|---|---|
| | Electron | Hole |
| GaAs *crystal*[a] | ~8000 | ~400 |
| Si *crystal*[a] | ~1500 | ~500 |
| CdTe *crystal*[a] | ≤~1000 | ≤~100 |
| PbTe[a] | ~6000 | ~4000 |
| PbS[a] | ~600 | ~700 |
| CIS *crystal*[b] | | ≤~300 |
| CZTSSe *crystal*[c] | 10-200 | ~1 |
| Organic *crystal*[d] | ≤~0.1 | ≤~15 |
| $CH_3NH_3SnI_3$ *crystal, pellet*[e] | | 200-300 |
| $CH_3NH_3PbI_3$ & $CH_3NH_3PbBr_3$ *crystals*[f] | ≤~100 | ≤~100 |

Data are based on:
[a] ref. 15
[b] ref. 51
[c] ref. 52 and ref. 53
[d] ref. 55
[e] ref. 49 and ref. 54
[f] refs. 5–7,37,38
*While singular measurements of very high mobilities in some perovskites have been reported (see text), here only independently reproduced values are given.

To be published in: *Journal of Physical Chemistry Letters,* Vol. 23, 2015; http://dx.doi.org/10.1021/acs.jpclett.5b02390
(link will be active from Dec. 3 2015)(40) Choi, J. J.; Yang, X.; Norman, Z. M.; Billinge, S. J. L.; Owen, J. S. Structure of Methylammonium Lead Iodide Within Mesoporous Titanium Dioxide: Active Material in High-Performance Perovskite Solar Cells. *Nano Lett.* **2014**, *14* (1), 127–133.

(41) He, Y.; Galli, G. Perovskites for Solar Thermoelectric Applications: A First Principle Study of $CH_3NH_3AI_3$ (A = Pb and Sn). *Chem. Mater.* **2014**, *26* (18), 5394–5400.

(42) Wang, Y.; Zhang, Y.; Zhang, P.; Zhang, W. High Intrinsic Carrier Mobility and Photon Absorption in the Perovskite $CH_3NH_3PbI_3$. *Phys Chem Chem Phys* **2015**, *17* (17), 11516–11520.

(43) Chen, Y.; Yi, H. T.; Wu, X.; Haroldson, R.; Gartstein, Y.; Zakhidov, A.; Zhu, X.-Y.; Podzorov, V. Ultra-long carrier lifetimes and diffusion lengths in hybrid perovskites revealed by steady-state Hall effect and photoconductivity measurements. *submitted for publication*.

(44) Juarez-Perez, E. J.; Sanchez, R. S.; Badia, L.; Garcia-Belmonte, G.; Kang, Y. S.; Mora-Sero, I.; Bisquert, J. Photoinduced Giant Dielectric Constant in Lead Halide Perovskite Solar Cells. *J. Phys. Chem. Lett.* **2014**, *5* (13), 2390–2394.

(45) Tretiak, S., *private communication*.

(46) Frost, J. M.; Butler, K. T.; Brivio, F.; Hendon, C. H.; van Schilfgaarde, M.; Walsh, A. Atomistic Origins of High-Performance in Hybrid Halide Perovskite Solar Cells. *Nano Lett.* **2014**, *14* (5), 2584.

(47) Liu, S.; Zheng, F.; Koocher, N. Z.; Takenaka, H.; Wang, F.; Rappe, A. M. Ferroelectric Domain Wall Induced Band Gap Reduction and Charge Separation in Organometal Halide Perovskites. *J. Phys. Chem. Lett.* **2015**, *6* (4), 693–699.

(48) Zheng, F.; Tan, L. Z.; Liu, S.; Rappe, A. M. Rashba Spin-Orbit Coupling Enhanced Carrier Lifetime in $CH_3NH_3PbI_3$. *Nano Lett.* **2015**, available online, DOI: 10.1021/acs.nanolett.5b01854.

(49) Stoumpos, C. C.; Malliakas, C. D.; Kanatzidis, M. G. Semiconducting Tin and Lead Iodide Perovskites with Organic Cations: Phase Transitions, High Mobilities, and Near-Infrared Photoluminescent Properties. *Inorg. Chem.* **2013**, *52* (15), 9019–9038.

(50) Stoumpos, C. C.; Malliakas, C. D.; Peters, J. A.; Liu, Z.; Sebastian, M.; Im, J.; Chasapis, T. C.; Wibowo, A. C.; Chung, D. Y.; Freeman, A. J.; et al. Crystal Growth of the Perovskite Semiconductor $CsPbBr_3$: A New Material for High-Energy Radiation Detection. *Cryst. Growth Des.* **2013**, *13* (7), 2722–2727.

(51) Neumann, H.; Nowak, E.; Kühn, G. Impurity States in CuInSe2. *Cryst. Res. Technol.* **1981**, *16* (12), 1369–1376.

(52) Gokmen, T.; Gunawan, O.; Mitzi, D. B. Minority carrier diffusion length extraction in $Cu_2ZnSn(Se,S)_4$ solar cells. *J. Appl. Phys.*. **2013**, *114,* 114511.

(53) Gunawan, O.; Virgus, Y.; Tai, K. F. A Parallel Dipole Line System. *Appl. Phys. Lett.* **2015**, *106* (6), 062407.

(54) Takahashi, Y.; Hasegawa, H.; Takahashi, Y.; Inabe, T. Hall Mobility in Tin Iodide Perovskite $CH_3NH_3SnI_3$: Evidence for a Doped Semiconductor. *J. Solid State Chem.* **2013**, *205*, 39–43.

(55) Podzorov, V. Organic Single Crystals: Addressing the Fundamentals of Organic Electronics. *MRS Bull.* **2013**, *38* (01), 15–24.
7